\documentclass[epj]{svjour}

\usepackage{graphicx}
\usepackage{epsfig}
\usepackage{fancyhdr}

\newcommand{\eg}{\textit{e.g.},}

\def\makeheadbox{{
 }}
\def\mytitle{My title}
\def\myauthors{My name}
\def\mytype{My type of session}
\def\mysession{My session}

\def\mytitle{Status of the EDELWEISS-2 Dark Matter Search} 
\def\myauthors{A.Chantelauze for the EDELWEISS collaboration}    
\def\mytype{Contributed Talk}
\def\mysession{Cosmology and Astrophysics}

\begin{document}
\title{Status of the EDELWEISS-2 Dark Matter Search}
\author{A.Chantelauze for the EDELWEISS collaboration\inst{1}
}                     
%
%
\institute{ Forschungszentrum Karlsruhe, Institut f\"{u}r Kernphysik, Postfach 3640, D-76021 Karlsruhe, Germany
} 
%
\date{}
\abstract{
The Edelweiss programme is dedicated to the direct search for Dark Matter as massive
weakly interacting particles (WIMPs) with Germanium cryogenic detectors operated in
the Laboratoire Souterrain de Modane in the French Alps at a depth of 4800 mwe.
After the initial phase Edelweiss I, which involved a total mass of 1 kg, the second
step of the programme, Edelweiss II, currently operates 9 kg of detectors and an active
shielding of 100 m$^2$ muon veto detectors and is now in its commissioning phase. The current
status and performance of the Edelweiss II set-up in terms of backgrounds will be given,
the underground muon flux measured with the muon veto system will be presented.
} 
\maketitle
%

\section{Introduction}
\label{intro}
Understanding the nature of Dark Matter in the universe is a major challenge for modern cosmology and astrophysics.
One of the well-motivated candidates is the generically named Weakly Interacting Massive Particle (WIMP).
In the Minimal Supersymetric Standard Model (MSSM) framework, the WIMP could be the Lightest Supersymmetric Particle, which is stable, neutral and massive.
The EDELWEISS (Experience pour DEtecter Les Wimps En Site Souterrain) experiment is dedicated to the direct detection of WIMPs trapped in the galactic halo. The detection is performed through the measurement of the recoil energy produced by elastic scattering of a WIMP off target nuclei.
The main challenges are the extremely low event rate of $\le$ 1 evt/kg/year and the relatively small deposited energy of $\le$ 100 keV.

\section{Experimental setup and detectors}
\label{sec:ExpSetup}

The EDELWEISS experiment is situated in the Laboratoire Souterrain de Modane (LSM) in the Fr\'{e}jus highway tunnel connecting Lyon with Turin below the French Alps.
The 1800 m of rock (4800 mwe) reduces the muon flux down to 4 $\mu$/m$^2$/d, a factor 10$^6$ times less than at the surface.
The EDELWEISS-II experiment installation was completed beginning of 2006.
EDELWEISS uses high purity Germanium cryogenic detectors, so-called bolometers, at a working temperature of T$\sim$20mK.
Such a configuration allows reading out an energy deposit in two distinct channels:
Heat via a small temperature rise and ionization via electrodes applying potential differences of a few Volts.
Whereas the heat channel reflects the total energy deposit, the amount of ionization strongly depends on the particle type which permits a strong suppression of electromagnetic radioactive background, as nuclear recoil produces less ionization in a crystal than an electron recoil does.
Therefore a discrimination parameter is defined as $Q=E$$_{ionization}$$/E$$_{recoil}$. If this parameter is normalized to 1 for electron recoils,  nuclear recoils, induced by WIMPs or neutron scattering, are observed at around $Q=0.3$, providing an excellent event-by-event discrimination.
However, nuclear recoils induced by neutrons will constitute the ultimate physical background which can mimic in its heat-to-ionization ratio a scattering process of a WIMP off a Germanium nucleus.
Neutrons therefore have to be discriminated by other experimental means.

\sloppy
The actual limitation of our present detectors arises from incomplete charge collection for near-surface events.
To reach cross-section sensitivities of interest for SUSY models, it is necessary to improve the rejection capabilities of the detectors in parallel with increasing the active mass.
Specific improvements are aimed at reducing the possible background sources \cite{Ref1} that have limited the sensitivity of EDELWEISS-I \cite{Ref2}.

\begin{figure}[t]
\centerline{\epsfxsize=3.2in\epsfbox{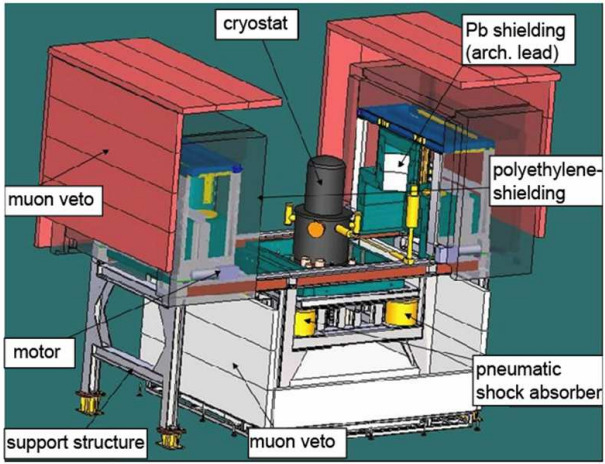}}
\caption{Schematic view of the EDELWEISS-II experiment. From outside to inside: The outer shell is the muon veto system, followed by the polyethylene shield and the inner lead shielding. The upper part can open to have access to the cryostat which houses the bolometers.}
\label{EDWSetup}
\end{figure}

\fussy
To reduce environmental background, all materials used in the vicinity of the detectors have been tested for their radiopurity with a dedicated HPGe detector.
A class 10,000 clean room surrounds the whole experiment and a class 100 laminar flow with deradonised air of $\le0.1$ Bq/m$^3$, is used to mount the detectors in the cryostat.
The gamma background is screened by a 20 cm thick lead shielding around the cryostat, see Fig. \ref{EDWSetup}.
Sources for neutrons are radioactivity in the rock and $^{238}$U fission in the lead shield as well as cosmic muons entering the detector environment.
In the experimental volume of the cryostat, the low energy neutron background in the cavity is attenuated by three orders of magnitude thanks to a 50 cm thick polyethylene shielding.
In addition a 100 m$^2$ plastic scintillator active muon veto surrounding the experiment with a coverage of $\ge$98$\%$ will tag the muons interacting in the lead shielding and producing neutrons.
The residual neutron background comes from high energy neutrons produced by muons that are not tagged by the veto system and by the neutrons from $^{238}$U fission in the lead shielding.
The nuclear recoil rate above 10 keV in the detectors is estimated by Monte Carlo simulation to be less than 10$^{-3}$
evt/kg/d, which is equivalent to a WIMP-nucleon cross-section sensitivity of 10$^{-8}$ pb for WIMP masses of $\sim$100 GeV/c$^2$.
This corresponds to an improvement of a factor 100 compared to EDELWEISS-I sensitivity.

The dilution cryostat used in the EDELWEISS-II experiment is of a reversed design, with the experimental chamber on the top of the structure.
It is a nitrogen-free system, using three pulse tubes to cool the 100 K and 20 K screens and a He reliquifier reducing the He consumption.
The experimental volume is about 50 liters allowing the installations of 110 detectors of about 330 g each in a compact arrangement allowing self shielding and multiple interactions identification.

EDELWEISS-II runs three different types of detectors :
(i) The known 330 g Ge/NTD type as used in EDELWEISS-I but equipped with new holders,
(ii) The 400 g Ge/NbSi type \cite{Ref3}\cite{Ref4} with two NbSi Anderson insulator thermometric layers for active surface events rejection that have been developed within the EDELWEISS collaboration,
(iii) A new type of detectors, the 400 g Ge/NTD crystals with a special interdigitized electrode scheme
recently developed \cite{Ref5}\cite{Ref6}, with the same aim to have active surface event rejection.
The surface event rejection capabilities of Ge/NbSi and Ge/NTD with interdigitized electrodes have been measured in surface laboratories to be better than 95$\%$ \cite{Ref4}\cite{Ref6} for events occurring in the first millimeters under the surface.

EDELWEISS-II has initially been funded for a 28 detectors stage:
21 Ge/NTD and 7 Ge/NbSi detectors.
Data acquisition with this setup started in the summer 2007.
Forty additional detectors (50$\%$ Ge/NbSi and 50$\%$ Ge/NTD with interdigitized electrodes) will be added in the two upcoming years to enhance progressively the sensitivity.

\section{Commissioning runs}
\label{sec:ComRun}

In comparison with EDELWEISS-I, EDELWEISS-II is a completely new experiment: New cryostat, up to 110 detectors, new electronics, new acquisition hardware and software and a fully numerical triggering system, which needed extensive tests, tuning and validation.

Four cryogenic runs with 8 detectors have been performed in 2006 to check and validate the different parts of the setup.
This whole year has been dedicated to the tuning of the electronics and improvements on the acquisition and cryogenics, especially for acoustic and mechanical decoupling of the thermal machines.
The quasi-total 28 detectors stage (21 Ge/NTD, 4 Ge/NbSi) has been mounted early 2007 for commissioning runs.

In addition to gamma and neutron source calibrations, low background runs have been performed to quantify the alpha, gamma, and beta backgrounds. The aim of these runs is to measure the rate of surface events with incomplete charge collection in the detectors and to extrapolate to the future sensitivity.

\begin{figure}[h]
\centerline{\epsfxsize=2.2in\epsfbox{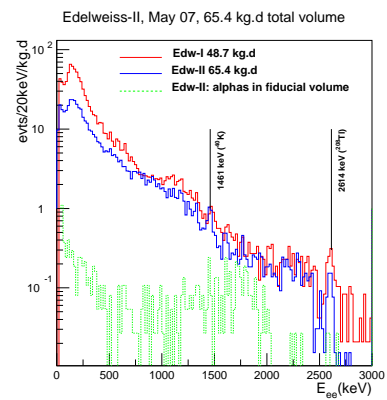}}
\caption{Ionization energy spectrum for low background run for the Ge/NTD detectors (65 kg.d for the total volume). The comparison with EDELWEISS-I gamma background shows an improvement of a factor 2 at low energy. The alpha background are identified by selecting events with a ionisation/recoil energy ratio less than 0.5.}
\label{fondltd}
\end{figure}

The electronic recoil background rate is shown in Fig. \ref{fondltd}.
For the fiducial volume \cite{Ref2}, this rate is around 0.6 evt/kg/d below 100 keV which is a factor 2.5 better than EDELWEISS-I.
The alpha background is between 1.6 and 4.4 $\alpha$/kg/d (75-200 $\alpha$/m$^2$/d) depending on the detector and its near-by environment.
The mean alpha rate for EDELWEISS-I was 4.2 $\alpha$/kg/d.

\begin{figure}[h]
\centerline{\epsfxsize=3.3in\epsfbox{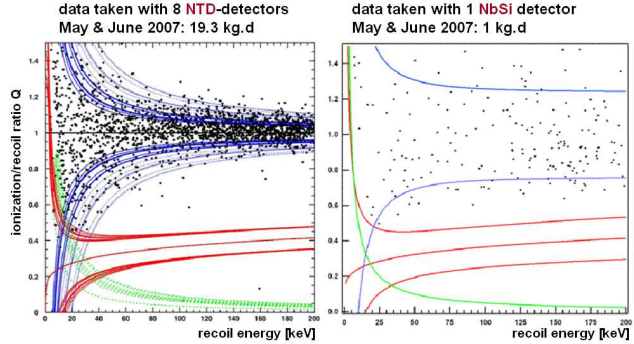}}
\caption{Quenching factor (ionisation/recoil energy ratio) vs. recoil energy in keV for background runs - Left hand side for 8 Ge/NTD detectors, right hand side for one Ge/NbSi detector.}
\label{results}
\end{figure}

Fig. \ref{results} shows the heat-to-ionization signal ratio as a function of the recoil energy for low energy background runs taken at the end of the commissioning period.
Results of this low background commissioning run corresponding to an exposure of 19.3 kg.d for the fiducial volume of the 8 Ge/NTD detectors with the lowest thresholds are shown on the left hand side.
The rate of events with incomplete charge collection, in the region of Q between 0.5 and 0.8 and measured with the method defined in \cite{Ref1} is comparable to EDELWEISS-I results.
No events are seen in the nuclear recoil band signal, showing a possible improvement compared to EDELWEISS-I. More statistics are obviously needed to draw firm conclusions.
Results obtained with a 200g Ge/NbSi detectors are shown on the right hand side for a fiducial exposure of 1 kg.d after the cuts removing the surface events.
The discrimination parameter is obtained with the athermal phonon signal which causes the rather large dispersion around $Q=1$.
The Ge/NbSi analysis is explained in details in \cite{Ref4}.

\section{Muon veto system}
\label{sec:muvet}

Though there is a reduction of the cosmic muon flux by the rock overburden of more than 10$^6$ compared to sea level,
neutrons produced in deep inelastic scattering (DIS) of cosmic muons are the most prominent background in the upcoming second generation experiments searching directly for Dark Matter.
The identification of muons in the vicinity of the Germanium detectors allows a significant suppression of this background source and hence an improvement in the experimental sensitivity.

The veto system consists of 42 plastic scintillator modules of 65 cm width, 5 cm thickness and lengths between 2 m and 4 m. The total surface is about 100 m$^2$ surrounding almost hermetically the outer polyethylene shielding of the cryostat, see Fig. \ref{EDWSetup}.
Each scintillator module is read out at both ends yielding 84 channels for the muon veto data acquisition system.
The electronics as well as the data read-out of the muon system are fully independent from the bolometer system but linked via a fiber connection transferring the overall experiment synchronization as well as some restricted information about muon hits.
The muon veto system is stable in terms of measuring conditions since July 2006.

\begin{figure}[h]
\centerline{\epsfxsize=3.3in\epsfbox{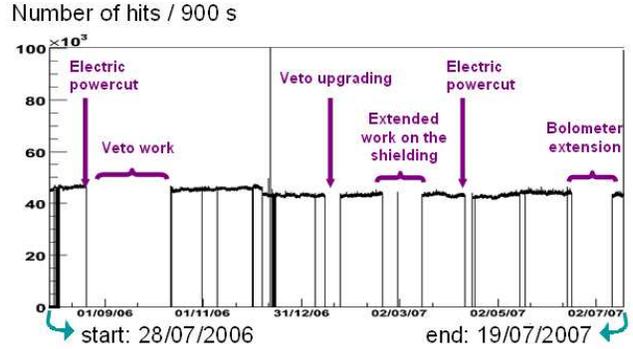}}
\caption{Example of a raw data plot for a side of module \#44, which can be translated into the lifetime of the veto system. During almost a year of stable conditions, the veto system was taking data for 60\% of the time.}
\label{VetoLifeTime}
\end{figure}

The raw data correspond to any signal collected by the photomultipliers at the end of a side of a module and above the threshold of the discriminator cards.
They are helpful to monitor the system.
With a rate of 8 kHz for the overall system, these are mainly composed of background.
Fig. \ref{VetoLifeTime} clearly summarizes the veto system live time: We have long periods with almost continuous data acquisition interrupted by electrical power cuts in the LSM laboratory as well as extensive work going on in the clean room of the experiment, when the high voltage of the veto system was switched off for safety reasons.
Above the rather constant raw data rate 
, there are periods with much higher event rates which could be identified as maintenance intervals of the cryostat/clean room in which the upper (mobile) veto modules were moved with the polyethylene shielding, or when radioactive sources were being manipulated.

The event data correspond to selected events, when there is at least one coincidence of both sides of a module in a 100 ns time window.
Requiring the event-defining coincidence, the hit rate of 8 kHz for raw data drops to 3 Hz for event data for the full muon veto system.
The event rate is still much higher than the expected muon rate of some $\mu$Hz for the full veto system \cite{RefA}\cite{Ref7} due to a deliberately low threshold not to miss any muons.
In an offline analysis, a set of candidates of throughgoing muons can be simply selected by, $\eg$ requiring a
coincidence between one module of the top side of the upper floor and any module of the bottom floor.


\begin{figure}[h]
\centerline{\epsfxsize=3.3in\epsfbox{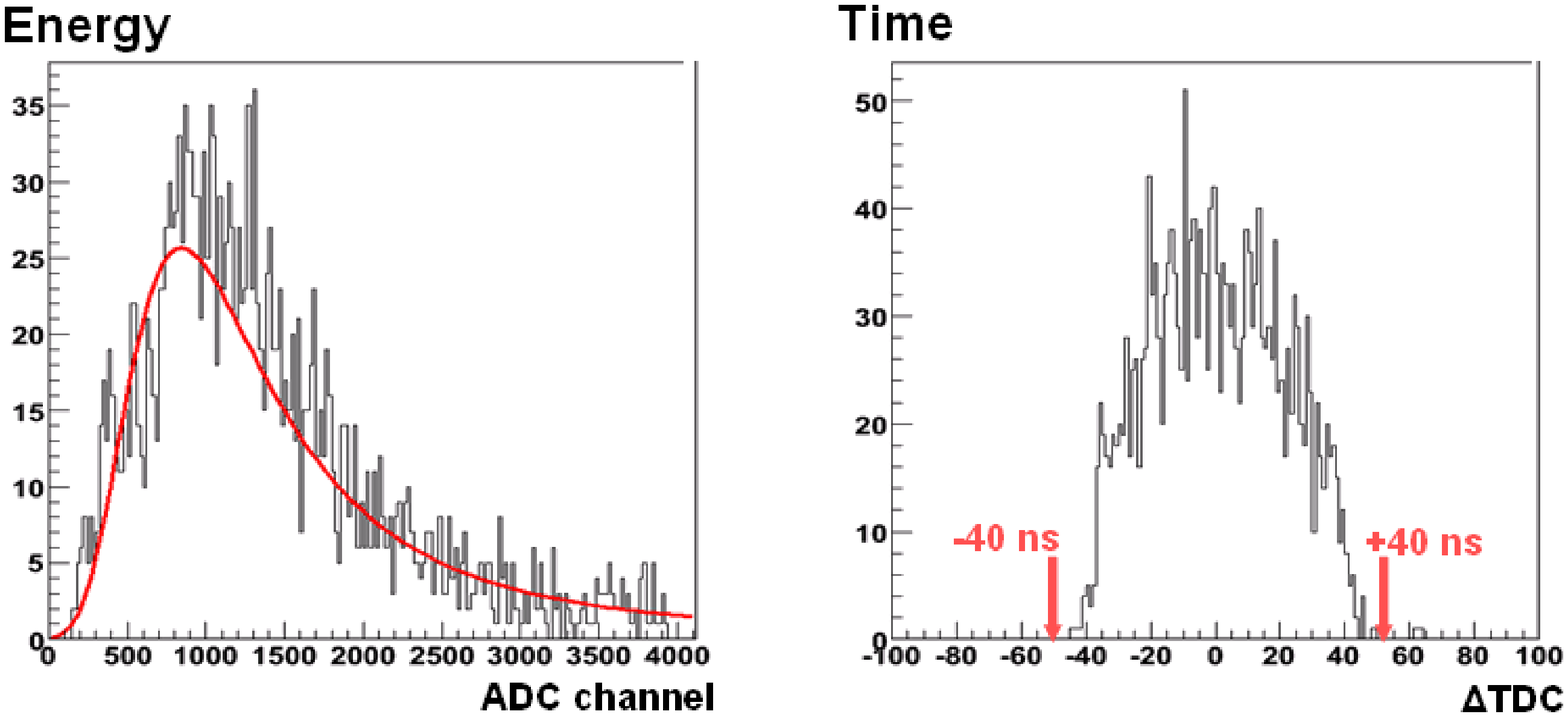}}
\caption{Example of data taken
from July 06 to July 07 (187 live days): Left Plot - Energy distribution of the muon candidates, the Landau fit is in red; Right plot - Time difference of both sides of a module, which corresponds to the position of the muon candidates along the axis of this module.}
\label{EnjTim}
\end{figure}

The relevant parameters for the data analysis are the ones related to the energy and the timing/position of the events.
Fig. \ref{EnjTim} shows the results for throughgoing muon candidates during a full year.
However, due to interruptions in the data acquisition, as shown in Fig. \ref{VetoLifeTime}, and to the selection of
a well-defined geometrical configuration (veto system closed),
which is the case in $\sim$70$\%$ of the running time, the effective measuring time for July 06 to July 07 corresponds to 187 live days.

On the left hand side of Fig. \ref{EnjTim} is the energy distribution of the muon candidates.
It is a preliminary plot, not yet calibrated in energy,
 with the Landau maximum
corresponding to a most probable energy of 10.8 MeV \cite{RefA}.
Even with the low statistics of $\sim$2000 muon candidates,
the distribution can be well fitted with a Landau distribution without requiring any background substraction.
On the right hand side of Fig. \ref{EnjTim}, the time difference of the signal coming from the two ends of a module is shown.
The time difference is due to the scintillation light path along the module axis
and can be translated to the position of the muon candidates along the axis of this module.

\begin{figure}[h]
\centerline{\epsfxsize=3.3in\epsfbox{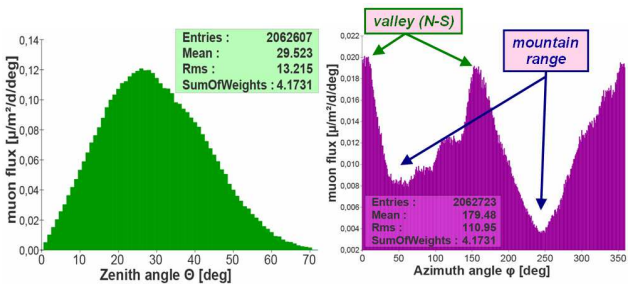}}
\caption{GEANT-4 detailed 3-D simulation of muons in
the LSM laboratory, including the full topology of the mountains \cite{RefA} - Left hand side, distribution of the zenith angle, right hand side, of the azimuth angle.}
\label{SimuShape}
\end{figure}

In parallel to the veto data analysis, a GEANT-4 detailed 3-D simulation of muons in
the LSM laboratory was performed, including the full topology of the mountains \cite{RefA}.
A first compilation of the rate of muon shows good agreement:
For the muon candidates shown in Fig. \ref{EnjTim}, the real data gives 11,2 $\pm$ 0,3 $\mu$/d
and the simulation 12,5 $\pm$ 0,4 $\mu$/d.
The simulation rate is normalized to earlier measurements from \cite{Ref7}.
This work is still preliminary,
the analysis period depending on the ongoing installation work and
the geometry of the shielding and the muon veto (upper part of the system open/closed).

The next step in the analysis of the event data of the muon veto system is to reconstruct the tracks of the muon candidates using the internal veto timing and to investigate coincidences of the muon veto system with bolometer events, to be expected at a rate of $\sim$1 event/week.

\section{Conclusions and prospects}
\label{sec:conc}

The EDELWEISS-II setup has been validated with calibration and background runs.
Energy resolutions and discrimination capabilities close to those of EDEL\-WEISS-I have been measured for Ge/NTD detectors.
Validation of Ge/NbSi detectors with new aluminium electrodes is in progress and Ge/NTD detectors with interdigitized electrodes scheme have shown promising results in surface laboratory.
The muon veto system runs, and study of coincidences between the muon veto and the bolometer systems will be performed in the following months.
A new setup for muon-neutron measurement will be soon installed at LSM and will allow to evaluate the muon induced neutron flux.
Low background physics runs will be taken with the 28 bolometers setup with the aim to reach sensitivity to WIMP-nucleon cross-section of $\sim$10$^{-7}$ pb for a WIMP mass of 100 GeV by July 2008.
40 additional detectors (Ge/NbSi and Ge/NTD with interdigitized electrodes) will be added in the two coming years to enhance progressively the sensitivity to $\sim$10$^{-8}$ pb thanks to active surface rejection, with an expected goal reached by autumn 2009/2010.

\sloppy
\begin{acknowledgement}
{\bf Acknowledgment:}
This work has been partially supported by the EEC Applied Cryodetector network (Contract HPRN-CT-2002-00322), the ILIAS integrating activity (Contract RII3-CT-2003-506222), the HGF initiative and networking fund through the virtual institute VIDMAN and the SFB/Transregio 27 of the Deutsche Forschungsgemeinschaft, DFG. The help of the technical staff of the Laboratoire Souterrain
de Modane and of the participating laboratories is gratefully acknowledged.
\end{acknowledgement}

%

\end{document}